\documentclass[aps, pra, superscriptaddress, reprint, floatfix, twocolumn,nofootinbib]{revtex4-2}

\usepackage{graphicx}

\usepackage{amsmath, amsthm, amssymb}
\usepackage{slashed}
\usepackage{bbold}
\usepackage[utf8]{inputenc}
\usepackage{empheq}
\usepackage{graphicx}
\usepackage{siunitx}
\usepackage[dvipsnames]{xcolor}
\definecolor{KB}{rgb}{0, 0.5, 0}

\usepackage[draft]{changes}
\definechangesauthor[name={Natalia}, color=magenta]{nat}
\definechangesauthor[name={Konstantin}, color=KB]{kb}
\definechangesauthor[name={Konstantin}, color=teal]{kbNew}
\definechangesauthor[name={Igor}, color=orange]{iv}
\definechangesauthor[name={Matteo}, color=blue]{mt}
\definechangesauthor[name={Christoph}, color=brown]{ck}

\usepackage{tikz}
\usetikzlibrary{shapes.geometric, arrows,angles}

\usepackage[plainpages=false, colorlinks=true, anchorcolor=blue, linkcolor=blue, citecolor=blue, bookmarks=false]{hyperref}

\newcommand{\bra}[1]{\left\langle #1 \right\rvert}
\newcommand{\ket}[1]{\left\lvert #1 \right\rangle}

\newcommand{\gs}{g^\phi}
\newcommand{\gp}{g^a}
\newcommand{\gv}{g^Z}
\newcommand{\gav}{g^B}
\newcommand{\rms}{r_\text{rms}}

\newcommand{\PsiN}{\mathrm{n}}
\newcommand{\PsiMu}{\mathrm{\mu}}
\newcommand{\PsiNBar}{\bar{\PsiN}}
\newcommand{\PsiMuBar}{\bar{\PsiMu}}

\begin{document}
\title{Challenging Beyond-the-Standard-Model Solutions to the Fine-Structure Anomaly in Heavy Muonic Atoms}

\author{K. A. Beyer}
\email[Author to whom correspondence should be addressed: ]{konstantin.beyer@mpi-hd.mpg.de}
\affiliation{Max-Planck-Institut f\"ur Kernphysik, Saupfercheckweg 1, 69117 Heidelberg, Germany}

\author{I. A. Valuev}
\affiliation{Max-Planck-Institut f\"ur Kernphysik, Saupfercheckweg 1, 69117 Heidelberg, Germany}

\author{C. H. Keitel}
\affiliation{Max-Planck-Institut f\"ur Kernphysik, Saupfercheckweg 1, 69117 Heidelberg, Germany}

\author{M. Tamburini}
\email{matteo.tamburini@mpi-hd.mpg.de}
\affiliation{Max-Planck-Institut f\"ur Kernphysik, Saupfercheckweg 1, 69117 Heidelberg, Germany}

\author{N. S. Oreshkina}
\email{natalia.oreshkina@mpi-hd.mpg.de}
\affiliation{Max-Planck-Institut f\"ur Kernphysik, Saupfercheckweg 1, 69117 Heidelberg, Germany}

\begin{abstract}
The leading-order contribution of a new boson to the muonic fine-structure anomaly, which refers to a discrepancy between the predicted transition energies and spectroscopic measurements of $\mu-^{90}$Zr, $\mu-^{120}$Sn, and $\mu-^{208}$Pb, is investigated. We consider bosons of scalar, vector, pseudoscalar, and pseudovector type. Spin-dependent couplings sourced by pseudoscalars or pseudovectors are disfavoured as solutions to the anomaly due to the nuclei in question having vanishing angular momentum. Spin-independent interactions resulting from scalar or vector exchange are also disfavoured because no parameter space exists to simultaneously fit different atomic states of the same nucleus. Therefore, we conclude that a `Beyond-the-Standard-Model' resolution of the muonic fine-structure anomaly is generally disfavoured, and the first-order solution by a single new boson is excluded. 

\end{abstract}

\maketitle

\section{Introduction}
Transition energies between bound muonic states in heavy atoms are of great interest for nuclear physics. The low-lying orbitals of muonic atoms have a sizeable overlap with the nucleus and are, as a result, highly sensitive to the nuclear structure \cite{Wheeler1949,Feinberg:1963xj,BorieRinker1982}. Spectroscopic measurements of the transition lines of $\mu-^{90}$Zr \cite{Phan:1985em}, $\mu-^{120}$Sn \cite{Piller:1990zza} and $\mu-^{208}$Pb \cite{Yamazaki:1979tf,Bergem:1988zz} reveal a discrepancy between theory prediction and experimental values, which we refer to as the fine-structure anomaly in heavy muonic atoms.

\begin{figure*}[t!]
    \centering
    \includegraphics[width=0.95\textwidth]{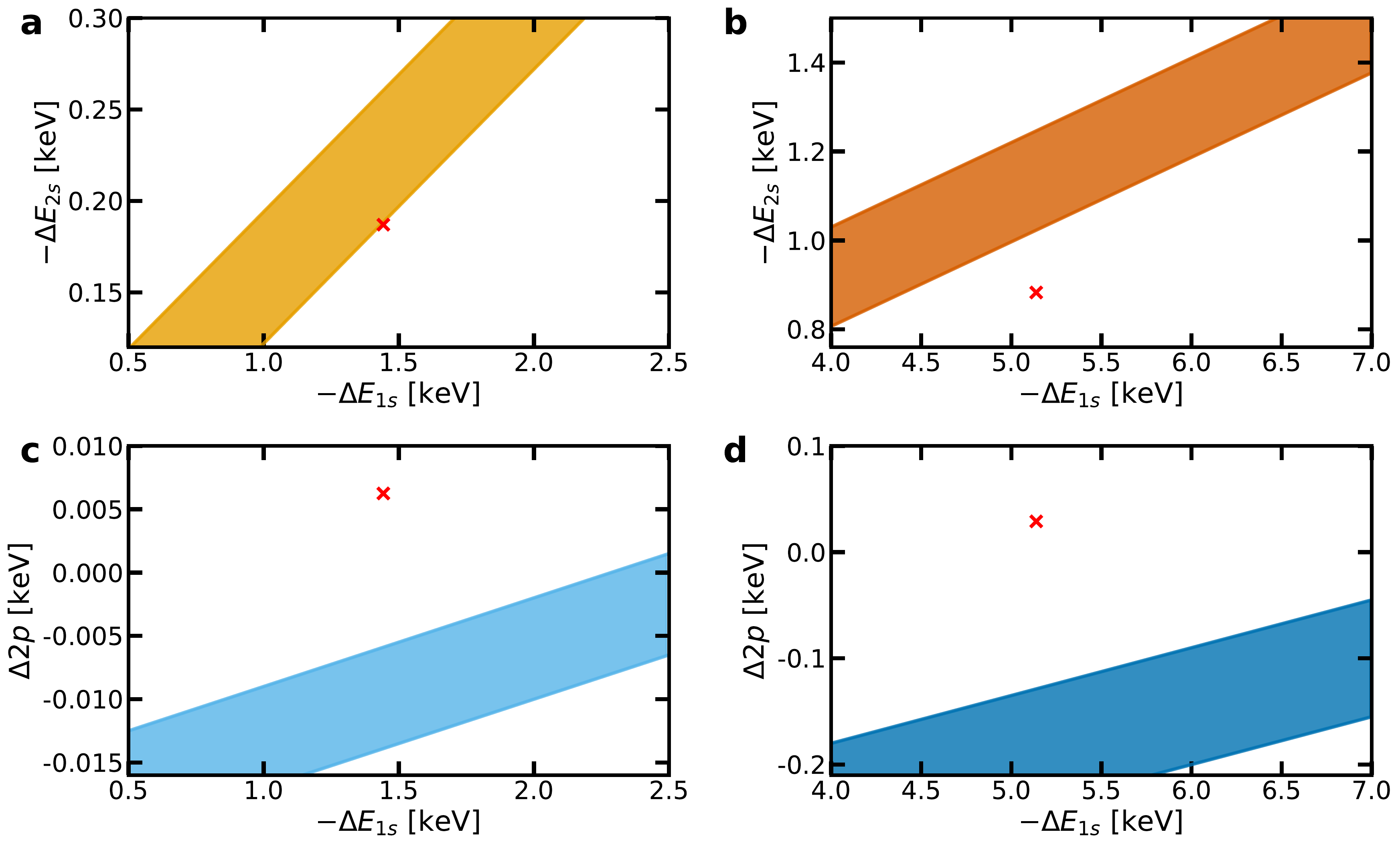}
    \caption{The coloured areas show the experimentally allowed ranges of shifts in $\Delta E_{2s}$ for $\mu-^{90}$Zr (a) and $\mu-^{208}$Pb (b), and $\Delta 2p$ for $\mu-^{90}$Zr (c) and $\mu-^{208}$Pb (d). The red crosses are the latest evaluations of the shifts from nuclear polarisation in the SAMi parametrisation as found in Ref.~\cite{Valuev:2022tau} and the self energy contributions as found in \cite{Oreshkina:2022mrk}. The plots are adapted from Ref.~\cite{Phan:1985em} for $\mu-^{90}$Zr and Ref.~\cite{Yamazaki:1979tf,Bergem:1988zz} for $\mu-^{208}$Pb, respectively. The slopes' parameters are given in table \ref{table:FigParam}.}
    \label{fig:ShiftsExp}
\end{figure*}

This puzzle was believed to be caused by nuclear polarisation effects; however, recent investigations revealed the corrections not sufficient to fit the experimental data \cite{Valuev:2022tau}. Theory prediction of the difference $\Delta 2p^\text{NP}\equiv \Delta E_{2p_{3/2}}^\text{NP} - \Delta E_{2p_{1/2}}^\text{NP}$ resulting from nuclear polarisation (NP) is consistently too high to be compatible with the precision measurements, see Fig.~\ref{fig:ShiftsExp}. Re-investigation of the original theory predictions is under way, see e.g.  Ref.~\cite{Oreshkina:2022mrk} for the muonic self-energy (SE) correction. Meanwhile, we carry out a complementary investigation of a potential `Beyond-the-Standard-Model' (BSM) solution to the muonic-atom fine-structure anomaly based on a similar line of argument: the heavier muon, bound on a tighter orbit around the nucleus, is more sensitive to new short-range forces than electrons on their larger orbits. The correction $\Delta E_{2s}^\text{NP}$ is approximately compatible with the experimental data and can therefore act as a reference: any new interaction should not spoil this compatibility.

New macroscopic forces, arising as a result of the exchange of new bosons added to the Standard Model (SM), come in a variety of forms \cite{Moody:1984ba,Dobrescu:2006au}. The nature of such new mediators dictates the phenomenology of the new forces. The inclusion of new scalars ($\phi$) or vectors~($Z_\nu$) coupled to the SM fermions $(f)$, to leading order, results in new, spin-independent forces. Pseudoscalars ($a$) and axial vectors ($B_\nu$) in contrast result in spin-spin couplings. The interaction of these new degrees of freedom is assumed to be of the following form
\begin{align}
\label{Eq:BSMInteractions}
	\mathcal{L}_\phi &\supset \gs_f \phi\bar{f}f,\qquad &&\mathcal{L}_a \supset i\gp_f  a\bar{f}\gamma^5 f,\nonumber\\
	\mathcal{L}_Z&\supset \gv_f Z_\nu\bar{f}\gamma^\nu f , &&\mathcal{L}_B\supset \gav_f  B_\nu\bar{f}\gamma^\nu\gamma^5 f
\end{align}
where the $g_f^{X}$ are the coupling constants of the new boson $X$ to the standard model fermions $f=p,n,\mu$, and $\gamma^\nu$ are the Dirac matrices. We work in units where $\hbar=c=1$. Also note that we exclusively reserve `$\mu$' to label the muon to avoid confusion with Lorentz indices.

The potential for discovering such forces in electronic atomic systems has recently been under investigation~\cite{Dzuba:2009kn,Dzuba:2017puc,Dzuba:2018anu}. Potential effects include radiative corrections to the energy levels of atoms \cite{Debierre:2021wam}, which become isotope-dependent for couplings to neutrons resulting in King plot non-linearities \cite{Flambaum:2017onb,Debierre:2022hco}. Spectroscopy of hydrogen-like ions is used to search for such features \cite{Frugiuele:2016rii,Debierre:2022ags}. Contributions to the hyperfine splitting of muonic hydrogen as a result of a novel coupling to the proton were investigated in~Ref.~\cite{Dorokhov:2017jcx}.  New scalar interactions have received great attention in the light of the Higgs boson interaction \cite{RevModPhys.54.67,Brax:2010gp}, including its influence on bound $g$ factors \cite{Sailer:2022azt}. A systematic investigation of the effects of various new bosons in muonic atoms was performed in Ref.~\cite{Okun:1973xn} with the treatment restricted to non-relativistic muon wavefunctions. The possibility of including new particles with an exclusive coupling to muons and their effect on $\mu-$H was investigated in Ref.~\cite{Karshenboim:2014tka} under the assumption of dark-photon-type models, giving the new boson a vector- or axial-vector-type interaction and a coupling to the electromagnetic current.

We restrict ourselves to tree-level couplings between the new boson and the standard-model fermions as in Eq.~\eqref{Eq:BSMInteractions}, which cause a shift in the binding energy of muonic atoms. Defining $\Delta 2p^X\equiv \Delta E_{2p_{3/2}}^X-\Delta E_{2p_{1/2}}^X$ and $\Delta 3p^X\equiv \Delta E_{3p_{3/2}}^X-\Delta E_{3p_{1/2}}^X$, in analogy to the nuclear-polarisation shift, the resulting fit to the experimental data can easily be performed. We use the full set of measurements available, which includes $\Delta E_{2s}$ shifts together with the above defined $\Delta 2p$ and $\Delta 3p$. We begin by reviewing the interaction potentials stemming from the new interactions in \S \ref{Sec:IntPot}. The full set of potentials can also be found in, e.g., Ref.~\cite{Fadeev:2018rfl,Frugiuele:2021bic}. We then calculate the resulting shifts in the binding energies of the low-lying muonic orbitals in \S \ref{Sec:ErgStateShift}. Fitting the experimental data for all measurements available leaves a number of interesting parameter ranges, which we calculate in \S \ref{Sec:MuonicAtomPuzzle}. We then examine the possibility of solving the muonic-atom fine-structure anomaly with a single new-boson inclusion to the SM.

\section{\label{Sec:IntPot}Interaction Potential}
A new exchange boson $X$ results in a new interaction term in the Hamiltonian, see appendix \ref{App:ErgShiftCalc} for more explanations. We will in the following refer to this term as the interaction potential.

\subsection{Scalar interaction}
Perhaps the simplest of the new interactions in Eq.~\eqref{Eq:BSMInteractions} and, as a result of that, the most studied is the interaction mediated by a new scalar boson $\phi$ corresponding to the first term
\begin{equation}
    \mathcal{L}_\phi\supset \sum_{f=p,n,\mu}\gs_f \phi \Bar{f}f.
\end{equation}
The exchange results in a potential between the muon $\mu$ and a single nucleon of the form \cite{Fadeev:2018rfl}
\begin{equation}
    V^{(f)}_\phi(r)=-\frac{\gs_f\gs_\mu}{4\pi}\gamma_0\int d^3 y\frac{e^{-m_\phi r}}{r}\bar{f}(\mathbf{y})f(\mathbf{y}),
\end{equation}
where $m_\phi$ is the mass of the new scalar and the $\gamma_0$ acts on the muonic wavefunction. The nucleon ($f=p,n$) is treated non-relativistically, which is justified by the slow motion in the nucleus. Thence, the nuclear potential, as seen by the muon, is obtained by a suitable average over the nuclear density 
\begin{equation}
    \rho_{(f)}(r)=\frac{3N_f}{4\pi R_0  ^3}\Theta\left(R_0 -r\right),
\end{equation}
which we assume to be a uniformly charged sphere with radius $R_0 = \sqrt{5/3}\rms$, where $\rms$ is the root-mean-square radius as found in \cite{Schopper:2004qco}. It reads 
\begin{align}
\label{Eq:ScalarPotential}
    V_\phi(r)&=-\alpha^\phi_f\gamma^0\int\frac{e^{-m_\phi\lvert\mathbf{r}-\mathbf{R}\rvert}}{\lvert\mathbf{r}-\mathbf{R}\rvert}\rho_{(f)}(\mathbf{R})d^3\mathbf{R}\nonumber\\
    &=\frac{-3N_f\alpha^\phi_f}{\left(m_\phi R_0 \right)^3}\left[\Theta(r-R_0)\mathcal{C}_1 + \Theta(R_0-r)\mathcal{C}_2\right]\gamma^0.
\end{align}
Here, $\alpha^\phi_f\equiv\gs_f\gs_\mu/4\pi$,
\begin{equation}
    \mathcal{C}_1\equiv\frac{e^{-m_\phi r}}{r}\left[m_\phi R_0\cosh\left(m_\phi R_0 \right)-\sinh\left(m_\phi R_0 \right)\right]
\end{equation}
and
\begin{equation}
    \mathcal{C}_2\equiv m_\phi - e^{-m_\phi R_0}\left(1+m_\phi R_0\right)\frac{\sinh\left(m_\phi r\right)}{r}.
\end{equation}

\subsection{Vector interaction}
A vector $Z_\nu$ mediated interaction also results in a spin-independent force to lowest order in the momentum exchange and corresponds to the third term in Eq.~\eqref{Eq:BSMInteractions}
\begin{equation}
    \mathcal{L}_V\supset \sum_{f=p,n,\mu}\gv_f \Bar{f}\slashed{Z}f.
\end{equation}
The potential sourced by a nucleon $f=n,p$ is obtained similarly to the scalar case as \cite{Fadeev:2018rfl}
\begin{equation}
    V^{(f)}_Z(r)=\alpha^Z_f \gamma_0\gamma_\alpha\int d^3y\bar{f}\gamma^\alpha f\frac{e^{-m_Z r}}{r}.
\end{equation}
The $\gamma_\alpha$ act on the muonic wavefunction. Non-relativistic treatment of the constituents of the nucleus results in $$\bar{f}\gamma^\alpha f\gamma_0\gamma_\alpha= j_{(f)}^\alpha(\mathbf{y})\gamma_0\gamma_\alpha\sim \rho_{(f)}(\mathbf{y}).$$
In analogy to the scalar case this then results in a potential
\begin{align}
    \label{Eq:VectorPotential}
    V_Z(r)=-V_{\phi}(r)\gamma^0.
\end{align}
Note that the two are not equal in spin space but, in the non-relativistic limit for the muon, reduce once again to the same Yukawa-like potential only differing by a sign.

\subsection{Pseudoscalar interaction}
The first spin-dependent interaction corresponds to the second term in Eq.~\eqref{Eq:BSMInteractions}
\begin{equation}
    \mathcal{L}_a\supset \sum_{f=p,n,\mu} i\gp_f a \Bar{f}\gamma^5 f,
\end{equation}
and is a result of pseudoscalar $a$ exchange. The potential is \cite{Fadeev:2018rfl}
\begin{equation}
    V^{(f)}_{a}(r)=4\pi\alpha_f^a \gamma^0\gamma^5\int d^3 y\int\frac{d^3q}{(2\pi)^3}\bar{f}\gamma^5f \frac{e^{-i\mathbf{q}\cdot(\mathbf{x}-\mathbf{y})}}{\mathbf{q}^2+m_{a}^2}.
\end{equation}
Taking the non-relativistic limit in the nuclear part of the interaction $$\int d^3y \bar{f}\gamma^5 f e^{i\mathbf{q}\cdot\mathbf{y}}\sim -\frac{\mathbf{s}_f\cdot\mathbf{q}}{m_f}$$ leads to the potential:
\begin{equation}
    \label{Eq:PseudoScalarPotential}
    V_{a}(r)=-i\alpha^{a}_f\gamma^0\gamma^5(\mathbf{s}_N\cdot\boldsymbol{\nabla})\frac{e^{-m_{a}r}}{m_f r}
\end{equation}
with $\mathbf{s}_{N}$ being the spin vector of the nucleus.

Note that here we are assuming the nucleus to be point-like, an assumption which will be unproblematic because of the vanishing nuclear spin for the elements in question.

\subsection{Axial vector interaction}
The other spin-dependent new force, arising from axial vector $B_\nu$ exchange (the last term in Eq.~\eqref{Eq:BSMInteractions}), has a potential \cite{Fadeev:2018rfl}
\begin{equation}
    V^{(f)}_{B}(r)=\alpha^B_f\gamma_0\gamma_\alpha\gamma^5 \int d^3y \bar{f}\gamma^\alpha\gamma^5 f\frac{e^{-m_B r}}{r}.
\end{equation}
In the limit of a non-relativistic, static nucleus,
\begin{equation}
\label{Eq:AxialVectorPotential}
    V^{(f)}_{B}(r)=-\alpha^B_f\left(i\gamma^5\frac{\mathbf{s_N}\cdot\boldsymbol{\nabla}}{m_f}  + \boldsymbol{\Sigma}\cdot\mathbf{s}_N\right) \frac{e^{-m_B r}}{r}.
\end{equation}
Here, $\boldsymbol{\Sigma}=\gamma_0\boldsymbol{\gamma}\gamma^5$.

\section{\label{Sec:ErgStateShift}Correction to the atomic energy states}

We will, in the following, work under the assumption that the new coupling $\alpha_X\equiv g^{X}_f g^{X}_\mu/4\pi$ be sufficiently small to allow treatment of the BSM effects as perturbations to the central Coulomb potential. Hence, the muon's wavefunction is calculated as a solution to Dirac's equation in the central Coulomb potential of the nucleus alone. The functional form of the Dirac spinors for state~$Q$ is then \cite{2006sham.book.....D}
\begin{equation}
\label{Eq:MuonWavefunctions}
  \Psi_{ Q }=\begin{pmatrix} g_Q(r) \chi_{\kappa_Q}^{m_Q}(\theta,\varphi)\\if_Q(r) \chi_{-\kappa_Q}^{m_Q}(\theta,\varphi) \end{pmatrix},
\end{equation}
with the bi-spinors
\begin{equation}
    \chi_\kappa^m(\theta,\varphi)=\begin{pmatrix} -\frac{\kappa}{\lvert\kappa\rvert}\sqrt{\frac{\kappa+\frac{1}{2}-m}{2\kappa+1}}\mathcal{Y}_{\lvert\kappa+\frac{1}{2}\rvert-\frac{1}{2}}^{ m-\frac{1}{2}}(\theta,\varphi)\\ \sqrt{\frac{\kappa+\frac{1}{2}+m}{2\kappa+1}}\mathcal{Y}_{\lvert\kappa+\frac{1}{2}\rvert-\frac{1}{2}}^{ m+\frac{1}{2}}(\theta,\varphi) \end{pmatrix}.
\end{equation}
Here, $\kappa_Q\equiv(-1)^{j+l+1/2}\left(j+\frac{1}{2}\right)$ is the relativistic angular quantum number, $j$ is the total angular momentum quantum number with $m$ the projection onto the quantisation axis, $l$ is the orbital momentum quantum number, and $\mathcal{Y}_a^b(\theta,\varphi)$ are normalised, complex spherical harmonics.

The BSM coupling then causes a perturbative shift in the binding energy of the states $Q$
\begin{equation}
    \Delta E_ Q ^{X}=\bra{Q} V_X(\mathbf{r})\ket{Q}= \int \Psi^\dagger_ Q (\mathbf{r}) V_X(\mathbf{r})\Psi_ Q (\mathbf{r})d^3\mathbf{r}.
\end{equation}

\subsection{\label{subsec:SpinIndep}Spin independent}
The exchange potential for scalars (Eq.~\eqref{Eq:ScalarPotential}) and vectors (Eq.~\eqref{Eq:VectorPotential}) have identical, trivial angular dependence. Therefore the angular part of the energy shift integral can be solved analytically
\begin{align}
\label{Eq:ScalarErgShift}
	 \Delta E_ Q ^{X} =\mp\alpha_X\frac{3N_f}{\left(m_\phi R_0 \right)^3}\Bigg[&\int_0^{R_0} d r r^2\mathcal{C}_1\left(g_Q^2(r)\mp f_Q^2(r)\right)\nonumber\\
	 &+\int_{R_0}^\infty d r r^2\mathcal{C}_2\left(g_Q^2(r)\mp f_Q^2(r)\right)\Bigg].
\end{align}
where the minus sign corresponds to the scalar interaction~[Eq.~\eqref{Eq:ScalarPotential}], and plus to the vector one~[Eq.~\eqref{Eq:VectorPotential}].

For a study of the general trends the far-field approximation and the corresponding expression 
\begin{equation}
   \lim_{R_0\rightarrow 0} V_s(r)=-\alpha_\phi N_f \gamma^0\frac{e^{-m_\phi r}}{r},
\end{equation}
are sufficient. This would result in a significantly simpler integral for the energy difference
\begin{equation}
\lim_{R_0\rightarrow 0}\Delta E^X_Q=\mp\alpha_X N_f \int dr r^2 \left(g_Q^2(r)\mp f_Q^2(r)\right)\frac{e^{-m_X r}}{r}.
\end{equation}
Once again, the minus (plus) corresponds to scalar (vector) exchange.\footnote{Note that the potential for scalar and vector boson exchange both reduce to the Yukawa potential in the non-relativistic limit. For relativistic muons they differ not only by the sign but also $\gamma^0$.} We, however, present calculations using the average over the nuclear density with the root-mean-square radii tabulated in Ref.~\cite{2013ADNDT..99...69A}. Quantitatively the results differ by up to $50\%$ for the $1s$ energy calculation but qualitatively the interpretation is unaffected.

\subsection{Spin dependent}
Because of the dependence on spin, more care must be taken when calculating the spin-dependent energy shifts. The individual angular momentum quantum numbers $I$ and $j$ of the nucleus and muon, respectively, are no longer good quantum numbers. Only the total angular momentum of the system $\mathbf{F}=\mathbf{I}+\mathbf{j}$ is. Because of our non-relativistic treatment of the nucleus, we use
\begin{equation}\label{Eq:NuclearAveragePoint}
     \sum_n\bra{IM} \mathbf{s}_{(n)}\ket{IM'} \equiv \bra{IM} \mathbf{S}\ket{IM'} = \bra{IM} \mathbf{I}\ket{IM'}.
\end{equation}
Then, the energy shift is evaluated as
\begin{align}
    &\Delta E_Q^{a}  =i\frac{\alpha_a}{2 m_f} \sum_{Mm}  \sum_{M'm'} C^{FM_F}_{IMjm}C^{FM_F}_{IM'jm'} 
    \nonumber\\ & \times  \bra{IM} \mathbf{I}\ket{IM'}\cdot\bra{jm}  \mathbf{r}\frac{e^{-m_a r}}{r^3}\left(1+m_a r\right)\gamma^5\ket{jm'}. 
\end{align}
Here we used
\begin{equation}
    \ket{ F M_F I j}=\sum_{M,m}C^{FM_F}_{IMjm}\ket{I M}\ket{j m}.
\end{equation}
$C^{FM_F}_{IMjm}$ are Clebsch-Gordon coefficients and $\ket{IM}$ and $\ket{jm}$ are the nuclear and muonic parts of the wavefunction, respectively. $M$($m$) is the projection of the total angular momentum $I$($j$) of the nucleus (muon) onto the $z$ axis. $F$ is the total angular momentum of the atom with the projection $M_F$.

After applying the Wigner-Eckhard theorem for the muonic part twice we are left with
\begin{equation}
    \Delta E_ Q ^{a}=-\alpha_a k_{(\mu)} [F(F+1)-I(I+1)-j(j+1)].
\end{equation}
Here we have defined
\begin{align}\label{Eq:Epsilon}
   k_{(\mu)} &= \int g_Q(r)f_Q(r)\frac{1+m_a r}{2m_f}e^{-m_a r} dr \\ 
    & \times  \int \cos(\theta)\left[\left(\chi_{-\kappa}^\frac{1}{2}\right)^\dagger\chi_{\kappa}^\frac{1}{2}  + \left(\chi_{\kappa}^\frac{1}{2}\right)^\dagger\chi_{-\kappa}^\frac{1}{2}\right] d^2\Omega. \nonumber
\end{align}
Treating the nucleus as a homogeneously charged sphere of radius $R_0$, we need to evaluate the radial integrals numerically, while the angular part can be obtained in closed form. The measurements have been performed for the magic nuclei $^{90}$Zr, $^{120}$Sn and the double magic $^{208}$Pb. These nuclei are spherical and have vanishing angular momentum $\langle\hat{I}\rangle=0$. 
Therefore, the spin dependent potentials do not affect the energy levels to leading order: $\Delta E_a^{a}=0$.

\section{\label{Sec:MuonicAtomPuzzle}Fit to the Muonic Atom Puzzle}
Fine-structure measurements of $\mu-^{90}$Zr \cite{Phan:1985em}, $\mu-^{120}$Sn \cite{Piller:1990zza} and $\mu-^{208}$Pb \cite{Yamazaki:1979tf,Bergem:1988zz} reveal a discrepancy between theory and experiment. The theoretical predictions of the SM contributions to $E_{2p}^\text{SM}\equiv E_{2p_{3/2}}^\text{SM}-E_{2p_{1/2}}^\text{SM}$ do not match the measured $E_{2p}^\text{exp}\equiv E_{2p_{3/2}}^\text{exp}-E_{2p_{1/2}}^\text{exp}$. A similar anomaly, albeit a weaker one, exists between the calculated $E_{2s}^\text{SM}$ and the measured $E_{2s}^\text{exp}$ for $\mu-^{208}$Pb. The above mentioned theory prediction includes the latest evaluations of NP \cite{Valuev:2022tau} and SE effects \cite{Oreshkina:2022mrk} as listed in table \ref{table:FigParam}.

\begin{figure*}
    \centering
    \includegraphics[width=0.95\textwidth]{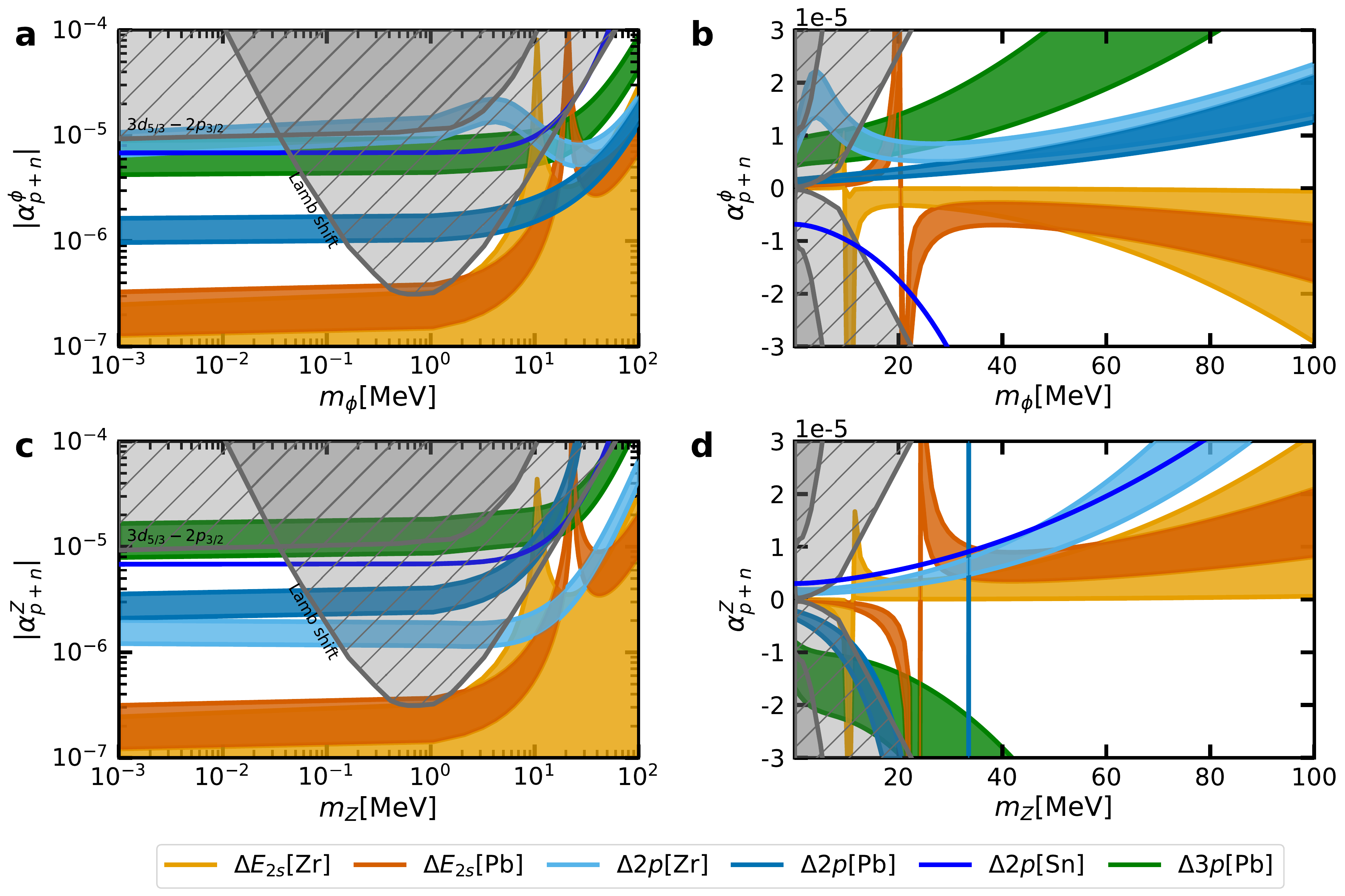}
    \caption{The plot shows the parameter space for a new scalar [vector] particle in logarithmic (a) [(c)] or absolute (b) [(d)] scales. The gray regions show previous exclusions coming from Lamb shift measurements of light muonic atoms and the transition $3d_{5/2}-2p_{3/2}$ in $\mu-^{24}$Mg and $\mu-^{28}$Si \cite{Beltrami:1985dc}. The lighter (darker) blue region shows the parameter space for which the level difference $\Delta 2p$ in $\mu-^{90}$Zr ($\mu-^{208}$Pb) fits the experimental measurements of Ref.~\cite{Piller:1990zza} (see also Refs.~\cite{Yamazaki:1979tf, Bergem:1988zz}). In yellow (orange) we depict the corresponding parameter space from fitting to the $\Delta E_{2s}$ states for $\mu-^{90}$Zr ($\mu-^{208}$Pb) and in green the $\Delta 3p$ measurements only available for $\mu-^{208}$Pb. In royalblue we show the fit for $\mu^{120}$Sn \cite{Piller:1990zza}.}
    \label{fig:ScalarCoupl}
\end{figure*}

The presence of a new force between the muon and the nucleus affects the binding energy and must be added to the SM prediction. Compatibility with the experimental values is obtained if
\begin{equation}
	E_Q^\text{SM}+\Delta Q^{X} = E_Q^\text{exp}.
\end{equation}
Fig.~\ref{fig:ShiftsExp} shows the required $\Delta 2p\equiv\Delta 2p^\text{NP}+\Delta 2p^{X}$ and $\Delta E_{2s}^\text{NP}+\Delta E_{2s}^{X}$ as coloured bands, and the latest NP + SE values as red crosses. Using this input we can calculate the coupling strength $\alpha_X$ necessary to fit the data for each nucleus (Zr, Sn, and Pb), and each state ($2s$, $2p$, and $3p$) individually. 
We parameterise the band's edges for the state $Q$ as
\begin{equation}
    \label{Eq:BandEquation}
    \Delta  Q (\Delta E_{1s})=-\mathcal{A}_ Q  \Delta E_{1s} + \mathcal{B}_Q,
\end{equation}
with the parameters as in table \ref{table:FigParam}.

\begin{table*}
\centering
\begin{tabular}{||c | c | c  | c  | c | r | r | r | r | r | r ||} 
 \hline
 Element &  $\mathcal{A}_{2p}^\text{Z}$ &  $\mathcal{A}_{2s}^\text{Z}$ & $\mathcal{B}_{2p,Z}^{(u,l)}$ [eV] & $\mathcal{B}_{2s,Z}^{(u,l)}$ [eV] & $\Delta E_{1s}^\text{NP}$[eV] & $\Delta E_{2s}^\text{NP}$[eV] & $\Delta 2p^\text{NP}$[eV] & $\Delta E_{1s}^\text{SE}$[eV] & $\Delta E_{2s}^\text{SE}$[eV] & $\Delta 2p^\text{SE}$[eV]\\ [0.5ex] 
 \hline 
 $^{90}$Zr &  $\SI{7e-3}{}$ & $0.15$ & $\SI{-16}{},\,\SI{-24}{}$ & $44,\,-28$ & $-1438$ & $-203$ & $5.7$ & $40$ & $16$ & $0.55$ \\[0.5ex] 
 $^{120}$Sn &  --- & --- & ---  & --- & $-2530$ & $-363$ & $18.7$ & $102$ & $-299$ & $-0.2$\phantom{0} \\[0.5ex]
 $^{208}$Pb &  $\SI{4.5e-2}{}$ & $0.19$ & $\SI{-360}{},\,\SI{-470}{}$ & $270,\,47$ & $-5727$ & $-1045$ & $59.1$ & $589$ & $162$ & $-31\phantom{.00}$ \\[0.5ex]
 \hline
\end{tabular}
\caption{A table showing the important parameters for the nuclei considered in the main text. The parameters for the compatibility region, parametrised as $\Delta  2p(\Delta E_{1s})=-\mathcal{A}_ {2p}  \Delta E_{1s} + \mathcal{B}_{2p}^{(u,l)}$ with $(u)$ [$(l)$] indicating the upper [lower] edge, were extracted from \cite{Phan:1985em} ($\mu-^{90}$Zr), \cite{Piller:1990zza}($\mu-^{120}$Sn) and \cite{Yamazaki:1979tf,Bergem:1988zz} ($\mu-^{208}$Pb). The parametrisation for $2s$ is equivalent. The NP shifts are calculated in \cite{Valuev:2022tau} and the SE corrections in \cite{Oreshkina:2022mrk}. Note that the SE correction quoted here is the difference between the previous evaluation and the modern value.} 
\label{table:FigParam}
\end{table*}

The above considered new-boson additions to the SM cause a shift in $\Delta E_{1s}$, $\Delta E_{2s}$ and $\Delta 2p$, which, for a fixed mass $m_X$, move the theory predictions on the lines
\begin{equation}
\label{Eq:NewPhys2pShift}
    \Delta Q=\Delta Q^\text{NP} + \left(\Delta E_{1s}-\Delta E_{1s}^\text{NP}\right)\left(\frac{\Delta Q^{X}}{\Delta E_{1s}^{X}}\right),
\end{equation}

NP data for $1s$ and $\Delta 2p$ have been presented in Ref.~\cite{Valuev:2022tau} and can be found in table \ref{table:FigParam}. Our evaluations for the $2s$ state using the SAMi parametrisation as an average of the alternative choices is also presented there. The latest evaluation of the SE contribution has been performed in Ref.~\cite{Oreshkina:2022mrk} and the differences to previous calculations are shown in table \ref{table:FigParam}.

The range of couplings which fit the experimental data can now simply be obtained by the intersection of the above curves Eq.~\eqref{Eq:NewPhys2pShift} with the corresponding experimental lines Eq.~\eqref{Eq:BandEquation}. The intersection points are
\begin{align}
    \alpha_X= \frac{\mathcal{B}_{2p}-\Delta 2p^\text{NP}-\Delta E_{1s}^\text{NP} \mathcal{A}_{2p}^Z}{\left[\left(\frac{\Delta 2p^{X}}{\alpha_X}\right)+\mathcal{A}_{2p}^Z\left(\frac{\Delta E_{1s}^{X}}{\alpha_X}\right)\right]},
\end{align}
from which we can easily infer the range by allowing any coupling between the two boundaries of the measurement range. The intersection for $\Delta E_{2s}$ is obtained analogously. Note that $\Delta 2p^{X}/\alpha_X$ and $\Delta E_{2s}^{X}/\alpha_X$ are independent of $\alpha_X$ and the sign of the required coupling is fixed.

Plots of the compatible ranges of couplings for scalars and vectors to all nucleons (protons and neutrons) with equal strength are given in Fig. \ref{fig:ScalarCoupl} a, b and c, d, respectively. Each coloured area in the plots indicates the necessary parameter range to fit the individual states. A simultaneous fit of multiple states and nuclei is therefore indicated by an intersection of the areas. A close inspection of the parameter space depicted in Fig. \ref{fig:ScalarCoupl} reveals that no such intersection of all the areas exists. Furthermore, even the fits for multiple states of the same element do not intersect, as can be seen when taking the sign requirement of the new coupling $\alpha^X$ into account. The required sign is determined by comparing the slope of the compatibility region $\mathcal{A}_Q$ with the slope of the line of new predictions $\Delta E_Q^{X}/\Delta E_{1s}^{X}$. The sign of the required coupling is inverted when the relative sizes of the slopes switch, corresponding to the poles in Fig. \ref{fig:ScalarCoupl}. For equal slopes, the required coupling formally diverges. This particularly disfavours a new scalar or vector as the solution to the muonic-atom-fine-structure anomaly because, while changing the ratio of couplings to different elements by lifting the assumption of equal coupling strength to all nucleons is possible, a state-dependent coupling can not be achieved in this way.

We are not restricted to couplings of these new bosons to protons and neutrons with equal strength only. In principle, there could be couplings to protons and neutrons with different strength and sign. To keep the number of new parameters to a minimum we restrict ourselves to a single new boson. The coupling to a specific nucleus depends on the new coupling to protons (neutrons) $\alpha^X_{p}$ ($\alpha^X_{n}$), both of which are identical for all nuclei, and the number of protons (neutrons) $Z$ ($A-Z$).  A simple substitution 
\begin{equation}
N_f\alpha_X = Z\alpha^X_{p} + (A-Z)\alpha^X_{n}
\end{equation}
is sufficient to transform the above calculations and plots. It is important to point out once more that the new coupling constant is independent of the atomic state $Q$ and therefore we will be unable to fit any single nucleus, which requires at least two different couplings for two different states. Comparison to Fig. \ref{fig:ScalarCoupl} shows that this is the case for both Zr and Pb. For most parameter space even the relative sign of the required coupling for different states is opposite. We must therefore conclude that a single new scalar or vector as in Eq.~\eqref{Eq:BSMInteractions} does not result in a successful resolution of the fine-structure anomaly in heavy muonic atoms.

\begin{figure*}
    \centering
    \includegraphics[width=0.95\textwidth]{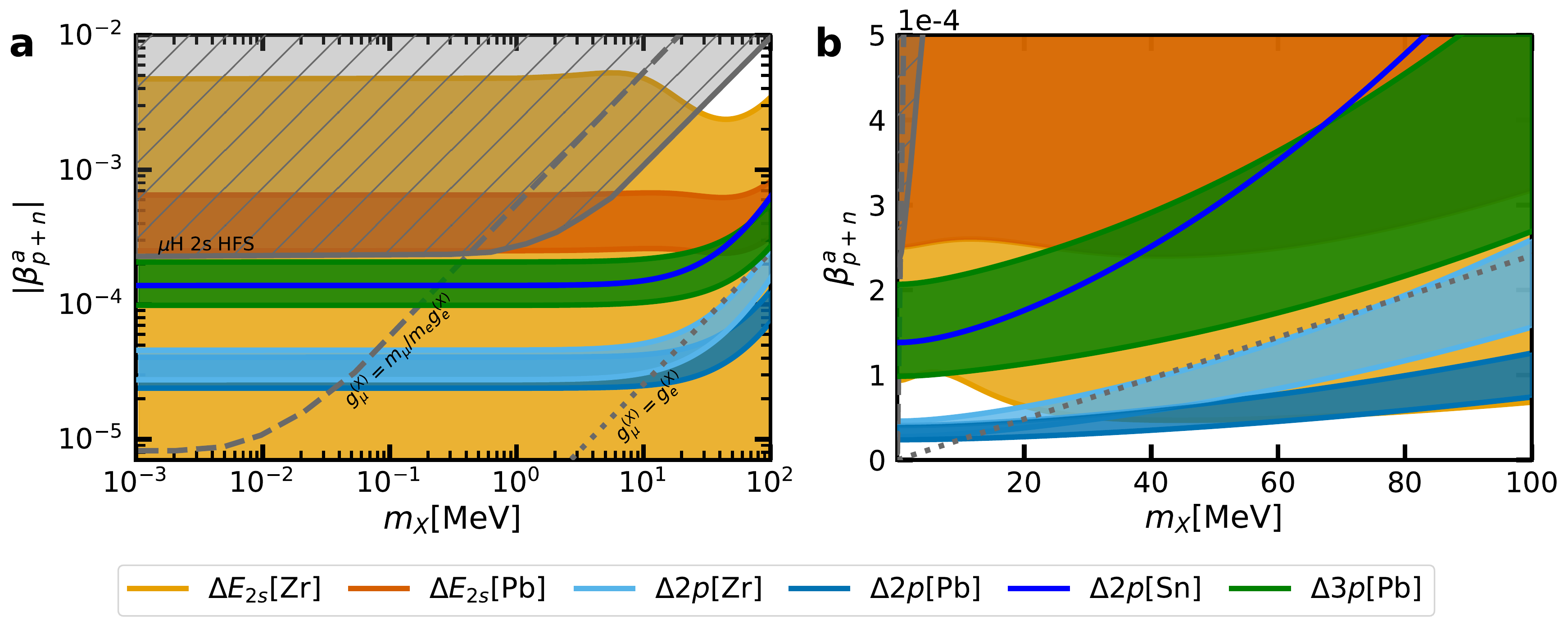}
    \caption{The plot shows the parameter space for a new pseudoscalar particle. The gray region shows previous exclusions coming from hyperfine measurements of muonic hydrogen \cite{Karshenboim:2010cg,Karshenboim:2010cj}. The other lines corresponds to projections of hyperfine measurements of electronic hydrogen under the assumption $g_\mu=m_\mu/m_e g_e$ (dashed) and $g_\mu=g_e$ (dotted).
The lighter (darker) blue region shows the parameter space for which the level difference $\Delta 2p$ in $\mu-^{90}$Zr ($\mu-^{208}$Pb) fits the experimental measurements of Ref.~\cite{Piller:1990zza} (\cite{Yamazaki:1979tf,Bergem:1988zz}). In yellow (orange) we depict the corresponding parameter space from fitting to the $\Delta E_{2s}$ states for $\mu-^{90}$Zr ($\mu-^{208}$Pb) and in green the $\Delta 3p$ measurements only available for $\mu-^{208}$Pb. In royalblue we show the fit for $\mu^{120}$Sn \cite{Piller:1990zza}.}
\label{fig:pseudoscalarCoupl}
\end{figure*}

A new pseudoscalar or axial vector degree of freedom will, to first order in the new couplings, not alleviate the tension because of the vanishing nuclear angular momentum of $\mu-^{90}$Zr, $\mu-^{120}$Sn and $\mu-^{208}$Pb. If we assume the nucleus to have angular momentum, for example by considering an excited state or an induced nuclear moment from the muon, we can easily perform the analogous analysis to the scalar case for the pseudoscalar and find Fig. \ref{fig:pseudoscalarCoupl} in which we have re-defined the coupling 
$$\beta_a\equiv\alpha_a \left(F(F+1)-I(I+1)-j(j+1)\right).$$ 
The three states $\Delta 2p[\text{Zr}]$, $\Delta 2p[\text{Pb}]$ and $\Delta E_{2s}[\text{Zr}]$ can be fit simultaneously for most pseudoscalar masses. The data for $\Delta 3p[\text{Pb}]$ and $\Delta E_{2p}[\text{Sn}]$ are not fit with the same boson. Both of these are problematic however, $\Delta 3p[\text{Pb}]$ calculations for NP being unreliable because of potential nuclear dipole excitations with energies close to the muonic $3p \rightarrow 1s$ transition \cite{BorieRinker1982,PhysRevA.39.5428,HAGA200571}, and the data for Sn do not come with a detailed analysis of the experimentally allowed NP corrections. The overlap is spoiled nevertheless by the fit to $\Delta E_{2s}[\text{Pb}]$.

A non-zero nuclear spin could arise, either because of excitation of the nucleus or as higher-order effects in the perturbation series. The former option may safely be excluded because the spectroscopic measurements reveal single transition lines instead of a forest of lines as would be expected for a mixture of excited and ground state nuclei. It is highly improbable that all nuclei are equally excited. The latter option is disfavoured because of the additional suppression of the effect, which would require the new coupling $\alpha_X$ to be larger, in conflict with $\mu$H measurements, as one can see from Fig. \ref{fig:pseudoscalarCoupl}. Additionally, any higher-order interactions would result in element-dependent shifts and therefore most likely spoil the nice overlap we see in the plot. We therefore conclude that the introduction of a new pseudoscalar or axial vector does not result in the potential resolution of the muonic fine-structure anomaly either.

Because of the sparseness of data any potential fit involving multiple new bosons is a result of the availability of more free parameters than data points to fit. We thus restrict ourselves to comment on the possibility but do not attempt such a fit until more data are available.

What can easily be seen is that two bosons of the same mass simply result in a re-definition of the coupling to $\alpha_X^{(1)}\pm\alpha_X^{(2)}$ and the above plots and conclusions apply to this combined coupling. If the nature of these two bosons of equal mass differs then the spin structure becomes $i\pm\gamma^5$ for a scalar with a pseudoscalar and $1\pm\gamma^5$ for a vector with a pseudovector. The contribution of these mixed potentials reduce to the above scalar or vector cases because the spin-spin coupling vanishes by the vanishing nuclear spin and the spin-density coupling is odd under parity, therefore does not result in a shift of the binding energy to first order in the new coupling. Hence, multiple new bosons of the same mass suffer the same problem as a single new boson.

\section{\label{Sec:Conclusion}Conclusion}
We have shown that the extension of the SM by one new boson of the form in Eq.~\eqref{Eq:BSMInteractions} will, to first order in the coupling, not result in the resolution of the fine-structure anomaly in heavy muonic atoms. We further argued against a solution involving combinations of multiple such bosons. While such a possibility is not excluded for more than one new boson with different masses, there does not exist enough data to meaningfully constrain the additional free parameters.

New spin-dependent forces as mediated by a pseudoscalar [Eq.~\eqref{Eq:PseudoScalarPotential}] or axial vector [Eq.~\eqref{Eq:AxialVectorPotential}] produce a shift of the binding energy proportional to the nuclear spin. In the case of $\mu-^{90}$Zr, $\mu-^{120}$Sn and $\mu-^{208}$Pb, the nuclear spin vanishes and therefore no change in the binding energy is produced to first order. Any potential higher-order contribution, like e.g. a muon induced static magnetic moment in the nucleus \cite{Beyer:2023}, will face challenges from previous exclusion bounds. As can be seen in Fig. \ref{fig:pseudoscalarCoupl}, the parameter range to fit the anomaly is close to the excluded region stemming from spectroscopy of $\mu$H. Higher-order effects will require larger couplings, scaling like the square root in the case of second-order effects in the new coupling.

A new scalar or vector boson results in a spin independent coupling which does produce a first-order shift in the binding energy. We have shown that there does not exist free parameter space for such a boson to simultaneously fit the spectroscopic data of $\mu-^{90}$Zr and $\mu-^{208}$Pb; the same nucleus requires different couplings $\alpha_X$ for different states. Fig. \ref{fig:ScalarCoupl} b, d reveal that even artificially increasing the error bars will not change this picture prior to the anomaly becoming statistically irrelevant.

It is therefore concluded that a BSM resolution of the fine-structure anomaly in heavy muonic atoms with a single new boson is disfavoured. This motivates further a careful re-investigation of the wide range of effects entering the binding energy calculations as was accomplished with NP in Ref.~\cite{Valuev:2022tau} or SE in Ref.~\cite{Oreshkina:2022mrk}. Finally, new experimental data including additional elements with improved precision could also shed light on the muonic puzzle, or even bring back some of the previously excluded resolutions.

\section*{Author's contributions}
K.A.B. did the calculations, made the figures, and wrote the manuscript. I.A.V. and  N.S.O. cross-checked the calculations. C.H.K. partook in discussions and gave comments on the manuscript.
M.T. initiated the project and lead the high-energy part. N.S.O. lead the atomic physics part. All authors participated in the discussions at all steps. 

\begin{acknowledgments}
We thank Zoltan Harman, Vladimir Zaytsev, Zewen Sun, and Giacomo Marocco for fruitful discussions and Laura Olivera-Nieto for help with the figures. 
\end{acknowledgments}
\bibliography{Bibliography}

\appendix
\widetext
\section{\label{App:ErgShiftCalc}Calculation of the Energy Shift}

\subsection{Scalar}
Be $\PsiN$, $\PsiMu$, and $\phi$ the nucleon, muon and exchange boson fields, respectively. The former two are Dirac 4 spinors and the latter is a bosonic scalar. The action for the coupled system is
\begin{equation}
	\mathcal{S}^s=\mathcal{S}_{(\mu)}+\mathcal{S}_{(n)}+\mathcal{S}_{(\phi)}+\mathcal{S}_{\gamma}+\mathcal{S}_{\text{int}}=\int \left(\mathcal{L}_{(\mu)}+\mathcal{L}_{(n)}+\mathcal{L}_{(\phi)}+\mathcal{L}_{(\gamma)}+\mathcal{L}_{\text{int}}\right)d^4x
\end{equation}
with the free particle Lagrangians
\begin{equation}
	\mathcal{L}_{(\mu)}+\mathcal{L}_{(n)}+\mathcal{L}_{(\phi)}^s +\mathcal{S}_{\gamma}=\PsiMuBar\left(i\slashed{D}-m_i\right)\PsiMu+\PsiNBar\left(i\slashed{D}-m_i\right)\PsiN + \frac{1}{2}\left(\partial_\nu \phi\right)^2 - \frac{1}{2}m_\phi^2\phi^2-\frac{1}{4}F^{\alpha\beta}F_{\alpha\beta},
\end{equation}
and the interaction Lagrangian
\begin{equation}
\mathcal{L}_{\text{int}}^s=g_\mu \PsiMuBar\PsiMu\phi + g_n \PsiNBar\PsiN\phi.
\end{equation}
Note that we have absorbed the electromagnetic interaction into the free particle Lagrangians for the electrically charged particles through the connection $D_\nu=\partial_\nu +iqA_\nu$.

The equations of motion (eom) for the fields are then, as obtained from the Euler-Lagrange equation
\begin{equation}
\label{Eq:EOMMuon}
	0=\frac{\delta\mathcal{L}_{{\mu}}}{\delta \PsiMu}-\frac{d}{dx^\nu}\frac{\delta\mathcal{L}_{(\mu)}}{\delta \partial_\nu\PsiMu}=\left(i\slashed{\partial}-q_{(\mu)}\slashed{A}-m_\mu+g_\mu\phi\right)\PsiMu ,
\end{equation}
\begin{equation}
	0=\left(i\slashed{\partial}-q_{(n)}\slashed{A}-m_n+g_n\phi\right)\PsiN ,
\end{equation}
and
\begin{equation}
\label{Eq:EOMScalarBoson}
	\left(\partial^2+m_\phi^2\right)\phi =g_\mu \PsiMuBar\PsiMu + g_n\PsiNBar\PsiN.
\end{equation}

We begin by making the assumption that $g_n\PsiNBar\PsiN\gg g_\mu \PsiMuBar\PsiMu$. In the heavy nuclei we wish to apply the analysis to, there are many nucleons in the nucleus. Making the assumption that $g_\mu\sim g_\mu$, it seems obvious that the main contribution to the static potential will come from the nucleus. Then Eq.~\eqref{Eq:EOMScalarBoson} has the solution
\begin{equation}
	\phi(x)=ig_n\int d^4y \int\frac{d^4q}{(2\pi)^4}\frac{ie^{iq.(x-y)}}{q^2-m_\phi^2}\PsiNBar(y)\PsiN(y).
\end{equation}
The equation for the muon Eq.~\eqref{Eq:EOMMuon} then becomes
\begin{equation}
	\left(i\slashed{\partial}-q_{(\mu)}\slashed{A}-m_\mu\right)\PsiMu(x)=4\pi\alpha_\phi \int d^4y \int\frac{d^4q}{(2\pi)^4}\frac{e^{iq.(x-y)}}{q^2-m_\phi^2}\PsiNBar(y)\PsiN(y)\PsiMu(x).
\end{equation}
From here we find
\begin{equation}
i\partial_t\PsiMu = \left(i\boldsymbol{\alpha}\cdot\boldsymbol{\nabla}+q_{(\mu)}A_0 - q_{(\mu)}\boldsymbol{\alpha}\cdot\mathbf{A}+m_\mu\gamma_0+4\pi\alpha_\phi \int d^4y \int\frac{d^4q}{(2\pi)^4}\frac{e^{iq.(x-y)}}{q^2-m_\phi^2}\PsiNBar(y)\PsiN(y)\gamma_0\right)\PsiMu\equiv \left(\mathcal{H}+\mathcal{H}_{(\phi)}^s\right)\PsiMu,
\end{equation}
where $\boldsymbol{\alpha}\equiv\gamma_0\boldsymbol{\gamma}$ and $\mathcal{H}$ is the Hamiltonian operator of the system. We further assume that the coupling to the new boson is small and does not significantly alter the muonic wavefunction, such that $\int d^3x\Psi_{Q}^\dagger \mathcal{H}\Psi_{Q}=E_{Q}$, the unperturbed binding energy of the muon. Then,
\begin{equation}
	\Delta E_Q^\phi=4\pi\alpha_\phi \int d^3 x \int d^4y \int\frac{d^4q}{(2\pi)^4}\frac{e^{iq.(x-y)}}{q^2-m_\phi^2}\PsiNBar(y)\PsiN(y)\mu^\dagger(x)\gamma_0\PsiMu(x).
\end{equation}

The Dirac bilinear in the limit of a static, non-relativistic nucleus reduces to $\PsiNBar(y)\PsiN(y)\sim \rho(\mathbf{y})$, the nuclear density. The energy difference then reduces to
\begin{equation}
	\Delta E_Q^\phi=-\alpha_\phi \int d^3 x \PsiMuBar(x)\PsiMu(x)\int d^3y\rho(\mathbf{y}) \frac{e^{-m_\phi r}}{r}.
\end{equation}

\subsection{Pseudo-scalar}
For $a$ a pseudoscalar, the free Lagrangian is unchanged but the interaction term changes to
\begin{equation}
	\mathcal{L}_{\text{int}}^p=ig_\mu \PsiMuBar\gamma^5\PsiMu a + ig_n \PsiNBar\gamma^5\PsiN a,
\end{equation}
to preserve CP invariance. Thence, the eom is
\begin{equation}
\label{Eq:EOMPseudoscalarBoson}
	\left(\partial^2+m_a^2\right)a =ig_\mu \PsiMuBar\gamma^5\PsiMu +i g_n\PsiNBar\gamma^5\PsiN,
\end{equation}
which is solved by
\begin{equation}
	a(x)=-g_n\int d^4y \int\frac{d^4p}{(2\pi)^4}\frac{ie^{ip.(x-y)}}{p^2-m_a^2}\PsiNBar(y)\gamma^5\PsiN(y).
\end{equation}
The muon equation is then
\begin{equation}
	\left(i\slashed{\partial}-q_{(\mu)}\slashed{A}-m_\mu\right)\PsiMu(x)=-4\pi\alpha_a \gamma^5 \int d^4y \int\frac{d^4q}{(2\pi)^4}\frac{e^{iq.(x-y)}}{q^2-m_a^2}\PsiNBar(y)\gamma^5\PsiN(y)\PsiMu(x),
\end{equation}
or the Hamiltonian
\begin{equation}
	i\partial_t\PsiMu = \left(i\boldsymbol{\alpha}\cdot\boldsymbol{\nabla}+q_{(\mu)}A_0 - q_{(\mu)}\boldsymbol{\alpha}\cdot\mathbf{A}+m_\mu\gamma_0-4\pi\alpha_a \gamma_0\gamma^5 \int d^4y \int\frac{d^4q}{(2\pi)^4}\frac{e^{iq.(x-y)}}{q^2-m_a^2}\PsiNBar(y)\gamma^5\PsiN(y)\right)\PsiMu\equiv \left(\mathcal{H}+\mathcal{H}_{(a)}^p\right)\PsiMu.
\end{equation}
The energy difference resulting from this perturbation is
\begin{equation}
	\Delta E_Q^a=-4\pi\alpha_\phi \int d^3 x \mu^\dagger(x)\gamma_0\gamma^5\PsiMu(x)\int d^4y \int\frac{d^4q}{(2\pi)^4}\frac{e^{iq.(x-y)}}{q^2-m_\phi^2}\PsiNBar(y)\gamma^5\PsiN(y).
\end{equation}
The Dirac bilinear for a static, non-relativistic nucleus is
\begin{align}
	\PsiNBar(\mathbf{y})\gamma^5\PsiN(\mathbf{y})\sim-\mathbf{s}\cdot\mathbf{q},
\end{align}
thence
\begin{equation}
	\Delta E_Q^a=i\alpha_a \int d^3 x \PsiMuBar(x)\gamma^5\PsiMu(x)\int d^3y \left(\mathbf{s}\cdot\boldsymbol{\nabla}\frac{e^{-m_a r}}{r}\right).
\end{equation}

\subsection{Vector}
Be $Z_\nu$ a vector boson. Then the free particle Lagrangian is
\begin{equation}
    \mathcal{L}_{Z}=-\frac{1}{4}\left(\partial^\alpha Z^\beta-\partial^\beta Z^\alpha\right)\left(\partial_\alpha Z_\beta-\partial_\beta Z_\alpha\right)+m_Z^2 Z_\nu Z^\nu\equiv-\frac{1}{4}F_{(Z)}^{\alpha\beta}F^{(Z)}_{\alpha\beta}+m_Z^2 Z_\nu Z^\nu,
\end{equation}
and the interaction term
\begin{equation}
    \mathcal{L}_\text{int}=g_\mu\PsiMuBar \slashed{Z}\PsiMu + g_n\PsiNBar\slashed{Z}\PsiN.
\end{equation}
Once again, we calculate the eom for the vector boson
\begin{equation}
    \label{Eq:EOMVectorBoson}
    \partial_\alpha\left(\partial^\alpha Z^\beta-\partial^\beta Z^\alpha \right)+ m_Z^2Z^\beta=\left(\partial_\alpha\partial^\alpha+m_Z^2\right)Z^\beta=g_n\PsiNBar\gamma^\beta\PsiN,
\end{equation}
where in the second equality we introduced the gauge condition $\partial_\alpha Z^\alpha=0$. The fundamental solution is
\begin{equation}
    D^{\alpha\beta}(q)=\frac{\left(\eta^{\alpha\beta}-\frac{q^\alpha q^\beta}{m_Z^2}\right)}{q^2-m_Z^2}
\end{equation}
and thus, the solution to the inhomogeneous equation is
\begin{equation}
    Z^\beta(x) = g_n \int d^4y \int\frac{d^4q}{(2\pi)^4} \frac{\left(\eta^{\alpha\beta}-\frac{q^\alpha q^\beta}{m_Z^2}\right)}{q^2-m_Z^2}e^{iq.(x-y)}\PsiNBar(y)\gamma_\alpha\PsiN(y)= g_n \int d^4y \int\frac{d^4q}{(2\pi)^4} \frac{e^{iq.(x-y)}}{q^2-m_Z^2}\PsiNBar(y)\gamma^\beta\PsiN(y).
\end{equation}
In the second equality we used the conservation of the Dirac current $\PsiNBar\gamma^\alpha\PsiN$. Once again, the equation for the muon becomes
\begin{equation}
    i\partial_t\PsiMu=\left(i\boldsymbol{\alpha}\cdot\boldsymbol{\nabla}+q_{(\mu)}A_0 - q_{(\mu)}\boldsymbol{\alpha}\cdot\mathbf{A}+m_\mu \gamma_0-\alpha_Z\gamma_0\gamma_\beta\int d^4y \int\frac{d^4q}{(2\pi)^4} \frac{e^{iq.(x-y)}}{q^2-m_Z^2}\PsiNBar(y)\gamma^\beta\PsiN(y)\right)\PsiMu\equiv \left(\mathcal{H}+\mathcal{H}_{(Z)}^V\right)\PsiMu.
\end{equation}
From here the energy difference is obtained as
\begin{equation}
	\Delta E_Q^Z = -\alpha_Z\int d^3 x \PsiMu^\dagger(x)\gamma_0\gamma_\beta\PsiMu(x)\int d^4y \int\frac{d^4q}{(2\pi)^4} \frac{e^{iq.(x-y)}}{q^2-m_Z^2}\PsiNBar(y)\gamma^\beta\PsiN(y) 
\end{equation}

The Dirac bilinear for a static, non-relativistic nucleus is $\PsiNBar(\mathbf{y})\gamma^\alpha\PsiN(\mathbf{y}) \sim j^\alpha(\mathbf{y})$, thence
\begin{align}
    \PsiMu^\dagger(x)\gamma_0\gamma_\beta\PsiMu(x)\PsiNBar(\mathbf{y})\gamma^\alpha\PsiN(\mathbf{y}) =j^0(\mathbf{y}) \PsiMu^\dagger(x)\PsiMu(x)-\mathbf{j}(\mathbf{y})\cdot \PsiMuBar(x)\boldsymbol{\gamma}\PsiMu(x)\sim\rho(\mathbf{y}) \PsiMu^\dagger(x)\PsiMu(x),
\end{align}
and the energy difference reduces to
\begin{equation}
	\Delta E_Q^Z=\alpha_Z \int d^3 x \PsiMu^\dagger(x)\PsiMu(x)\int d^3y \rho(\mathbf{y}) \frac{e^{-m_Z r}}{r}.
\end{equation}

\subsection{Pseudo-vector}
For $B^\nu$ a pseudo-vector, the calculation is equivalent to the vector case up to the Dirac bilinear
\begin{equation}
 \PsiNBar(\mathbf{y})\gamma_\nu\gamma^5\PsiN(\mathbf{y})\sim\begin{pmatrix}\frac{i}{m_\mu}\mathbf{s}\cdot\boldsymbol{\nabla} \\ \mathbf{s}\end{pmatrix}.
\end{equation}
Thence,
\begin{equation}
	\Delta E_Q^B=\alpha_B \int d^3 x \PsiMuBar(x)\gamma_\alpha\gamma^5\PsiMu(x)\int d^3y \begin{pmatrix}\frac{i}{m_\mu}\mathbf{s}\cdot\boldsymbol{\nabla} \\ \mathbf{s}\end{pmatrix}^{\alpha} \frac{e^{-m_Z r}}{r}.
\end{equation}

\end{document}